\pdfoutput=1

\documentclass[aps,prb,twocolumn,showpacs,superscriptaddress]{revtex4-1}

\usepackage{graphicx}
\usepackage{graphics}
\usepackage{amsmath}
\usepackage{amssymb}
\usepackage{amsfonts}
\usepackage{dcolumn}
\usepackage{dsfont}
\usepackage{latexsym}
\usepackage{rotating}
\usepackage{color}
\usepackage{latexsym}
\usepackage{bbm}
\usepackage{subfigure}
\usepackage{float}
\usepackage{epsfig}
\usepackage{psfrag}
\usepackage{bm}
\usepackage{amsthm}
\usepackage{eucal}
\usepackage{mathrsfs}
\usepackage{url}

\usepackage{color} 

\usepackage{hyperref}
\hypersetup{
colorlinks=true,final=true,
        linkcolor=blue,
        citecolor=blue,
        filecolor=blue,
        urlcolor=blue,
}
\begin{document}
\title{Real-time observation of spin-resolved plexcitons between metal plasmons and excitons of WS$_2$}
\author{R. K. Chowdhury}
\affiliation{Department of Physics, Indian Institute of Technology Kharagpur, Kharagpur - 721302, India}
\author{P. K. Datta}
\author{S. N. B. Bhaktha}
\affiliation{Department of Physics, Indian Institute of Technology Kharagpur, Kharagpur - 721302, India}
\author{S. K. Ray*}
\affiliation{Department of Physics, Indian Institute of Technology Kharagpur, Kharagpur - 721302, India}
\affiliation{S. N. Bose National Centre for Basic Sciences, Kolkata - 700106, India}

\begin{abstract}
\noindent
Strong light-matter interactions between resonantly coupled metal plasmons and spin decoupled bright excitons from two dimensional (2D) transition metal dichalcogenides (TMDs) can produce discrete spin-resolved exciton-plasmon polariton (plexciton). A few efforts have been made to perceive the spin induced exciton-polaritons in nanocavities at cryogenic conditions, however, successful realization of spin-resolved plexciton in time-domain is still lacking. Here, we are able to identify both the spin-resolved plexcitons discretely at room temperature and investigate their ultrafast temporal dynamics in size-tunable $Au-WS_{2}$ hybrid nanostructures using femtosecond pump-probe spectroscopy technique. Furthermore, we attribute that zero detuning between the excitons and plasmons is achieved at $\sim$7.0 ps along with transient Rabi-splitting energy exceeding $\sim$250 meV for both the spin-plexcitons, validating the strong-coupling conditions of polariton formation. Realization of these novel spin-plexcitons in the metal-TMDs platform is, therefore, interesting for both fundamental understanding and their possible futuristic applications in quantum photonics operating at room temperature.
  
\end{abstract}

\maketitle
\section{Introduction}
\noindent
Hybrid nanostructures interconnecting fundamental light-matter interactions of dissimilar components provide an innovative platform for designing futuristic photonic devices like polariton nanolasers~\cite{r1, r2}, spin-switches~\cite{r3}, digital data storage for quantum computing~\cite{r4}, single photon transistor~\cite{r5} and metamaterials~\cite{r6}. Even though practical realization of exciton-plasmon polariton devices is still underdeveloped, interesting experimental outcomes have been reported recently on plexcitonic hybrids composed of metallic nanostructures and cavities interacting with molecular excitons in organic semiconductors and dyes~\cite{r7, r8, r9, r10, r11, r12, nr1, newref1, newref2} or semiconducting excitons in 2D transition metal dichalcogenides (TMDs)~\cite{nr3, r13, r14, r15, nref1, nrf1}. Indeed, the superior coupling between optical field and surface-plasmons makes the metal nanostructures appealing for strong light-matter interactions~\cite{r16, r17}. On the other hand, excitons in TMDs ($MoS_2$, $WS_2$) are potentially attractive due to their massive exciton binding energy ($\sim$0.5 eV) and spin-orbit coupled photon energy selectiveness resulting in alike but independent spin-locked bright excitons ($X^0_A$ and $X^0_B$) in the same system~\cite{r18}. Thus, the coupling between spin resolved excitons and plasmons (P) can offer unexpected properties depending upon their coupling-strength. In general, surface-enhanced phenomena like Purcell effect, Fano resonances are dominant in the weak to intermediate-coupling regime~\cite{r19, r20}, whereas the strong-coupling occurs at ultrafast timescale (faster than electron relaxation) which can lead to the formation of spin-plexcitons ($X^0_A-P$ and $X^0_B-P$) in metal-TMDs hybrid nanostructures~\cite{r11, r23, nr5}, hitherto unexplored.

Interestingly, excitons in monolayer TMDs are highly confined along in-plane directions, whereas metal nanostructures usually trap the optical field in perpendicular to the layer planes~\cite{r21}. This misalignment of effective dipole-dipole interactions lead to poor coupling between $X^0$ and $P$. Therefore, the strong coupling can only be achieved, if more dipoles in TMDs are oriented along out-of-plane direction. A few-layer TMDs is a possible pathway instead of a monolayer to achieve stronger coupling, as proposed by Kleemann et al.~\cite{r21}, which eventually lead to the generation of plexcitons. In general, plexcitons are tracked down by observing the anti-crossing behavior of polariton dispersion curves and formation of two energetically well-separated hybridized peaks governed by Rabi-splitting, a measure of coupling-strength~\cite{r22}. Indeed, several reports show that plexcitonic states can be achieved by the fine tuning of oscillator strength and spectral linewidth of plasmons and excitons, via linear and angle-resolved spectroscopy techniques that required ultrahigh optical pumping ($\geq$10$^3$ W/cm$^2$)~\cite{r9, r10, nr2, nr4}. Alternatively, the ongoing quest is to tune exciton-plasmon overlap via two-step pump-probe spectroscopy technique that can separately create exciton ($X^0$) and plasmon ($P$) in a time-resolved process, resulting in the generation of plexcitons~\cite{r11, r23, nr5}.

Here, we report individual ultrafast detection of both the spin-resolved plexcitons ($X^0_A-P$ and $X^0_B-P$) by controlling the dimensions ($\sim$30 and 90 nm) of self-assembled gold nanoislands on layered $WS_2$ ($AuNI-WS_2$) fabricated on different substrates, such that their plasmonic resonance perfectly overlaps with individual spin-resolved excitons of $WS_2$. Moreover, a strong out-of-plane dipolar interaction is attributed between pump-induced excitons and probe-induced plasmons in time-domain. It results in remarkable transient Rabi-splitting energies as high as $\Omega_R$($X^0_A-P$) $\sim$250 meV and $\Omega_R$($X^0_B-P$) $\sim$270 meV, thus providing new insights of these novel spin induced plexcitonic hybrids.

\section{Results and discussion}
\label{result}
\noindent
\subsection{Individual overlap of excitons and plasmons}
\begin{figure}[t]
\centering
\begin{tabular}{cc}
\includegraphics[width=85mm]{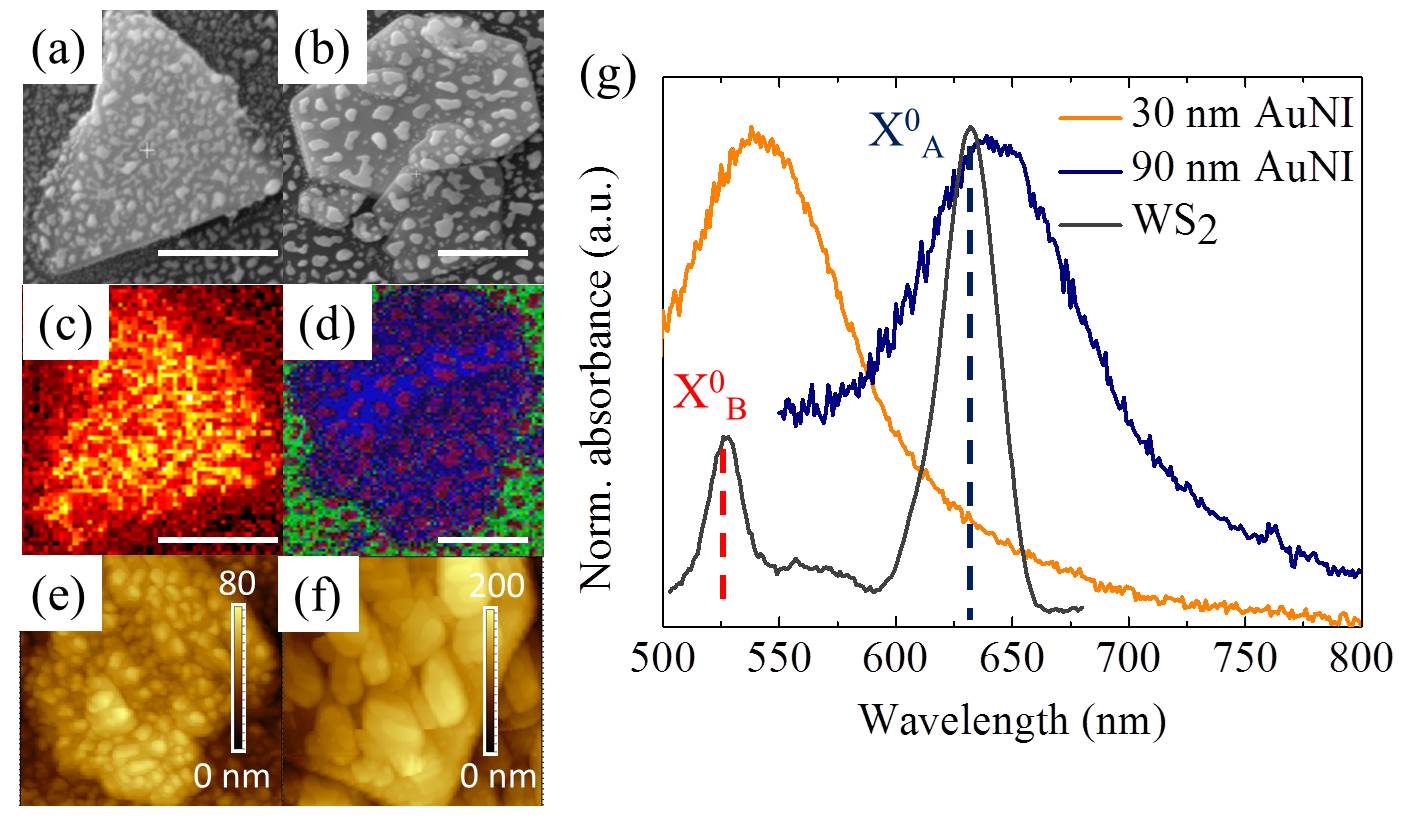}
\end{tabular}
\caption{(Color online)  Scanning electron micrographs of $Au$ nanoislands on layered $WS_2$ resulting in $AuNI-WS_2$ hybrid nanostructures for (a) $\sim$30 nm and (b) $\sim$90 nm diameter $Au$ nanoislands and (c and d) Auger electron mapping images of the corresponding hybrid samples showing compositional homogeneity. Here, yellow color in figure (c) and blue, red and green colors in figure (d) denote $S$, $W$, $Au$ and $Si$ elements, respectively confirming a uniform composition of samples. Here, scale bar is 500 nm for figure (a and b). (e and f) are the atomic force micrographs of $\sim$30 and $\sim$90 nm $AuNI-WS_2$ to confirm the presence of nanoislands on $WS_2$ layers. The root mean square surface roughness values are 8.8 nm and 23.8 nm for $\sim$30 and $\sim$90 nm $AuNI-WS_2$ samples, respectively. (g) Steady-state absorption spectra of $AuNI$ ($\sim$30 and 90 nm diameter) and bare $WS_2$ film on the glass substrate, where both the excitons of $WS_2$ individually match with the plasmonic modes of $AuNI$ (blue dashed line for $X^0_A$ and red dashed line for $X^0_B$) for different sizes.}
\label{fig1}
\end{figure}

To provide independent experimental evidence of both the spin-resolved plexcitons, we have fabricated scalable prototypes of $AuNI-WS_2$ hybrid nanostructures on glass and $Si$ substrates with controlled AuNI dimensions of $\sim$30 and 90 nm (see details in Materials and Methods). Scanning electron micrographs and atomic force micrographs confirm the size-distribution, roughness and surface profile of $Au$ nanoislands on $WS_2$ nanosheets, whereas Auger electron spectroscopy maps reveal the compositional homogeneity of the fabricated hybrids as shown in Fig.~\ref{fig1}(a)-(f) (also see Fig. S1, Supplementary Information). Here, we have chosen $Au$ nanostructures, a well-known non-reactive stable plasmonic system with efficient and wide spectral tunability (500-700 nm) in the desired wavelength regime by controlling their size and shape. To detect both the spin-plexcitons independently, we have optimized the size of self-assembled $Au$ nanoislands to be of $\sim$90 and $\sim$30 nm in diameter, because of their characteristic plasmon resonance wavelength coinciding individually with the respective spin-coupled bright excitons ($X^0_A$ and $X^0_B$) of $WS_2$, as shown in the steady state optical absorption spectra (Fig.~\ref{fig1}(g)). It is important to mention that we measured the steady-state spectra using a PerkinElmer spectrometer equipped with a halogen lamp as the excitation source of broadband UV-Visible illumination ($\sim$0.1 W/cm$^2$). The intensity is thus insufficient to generate exciton-plasmon polaritons (plexcitons) in the $AuNI-WS_2$ hybrid system using our steady-state $UV-vis$ absorption set-up~\cite{nr2, nr4}. 

\begin{figure}[t]
\centering
\begin{tabular}{cc}
\includegraphics[width=85mm]{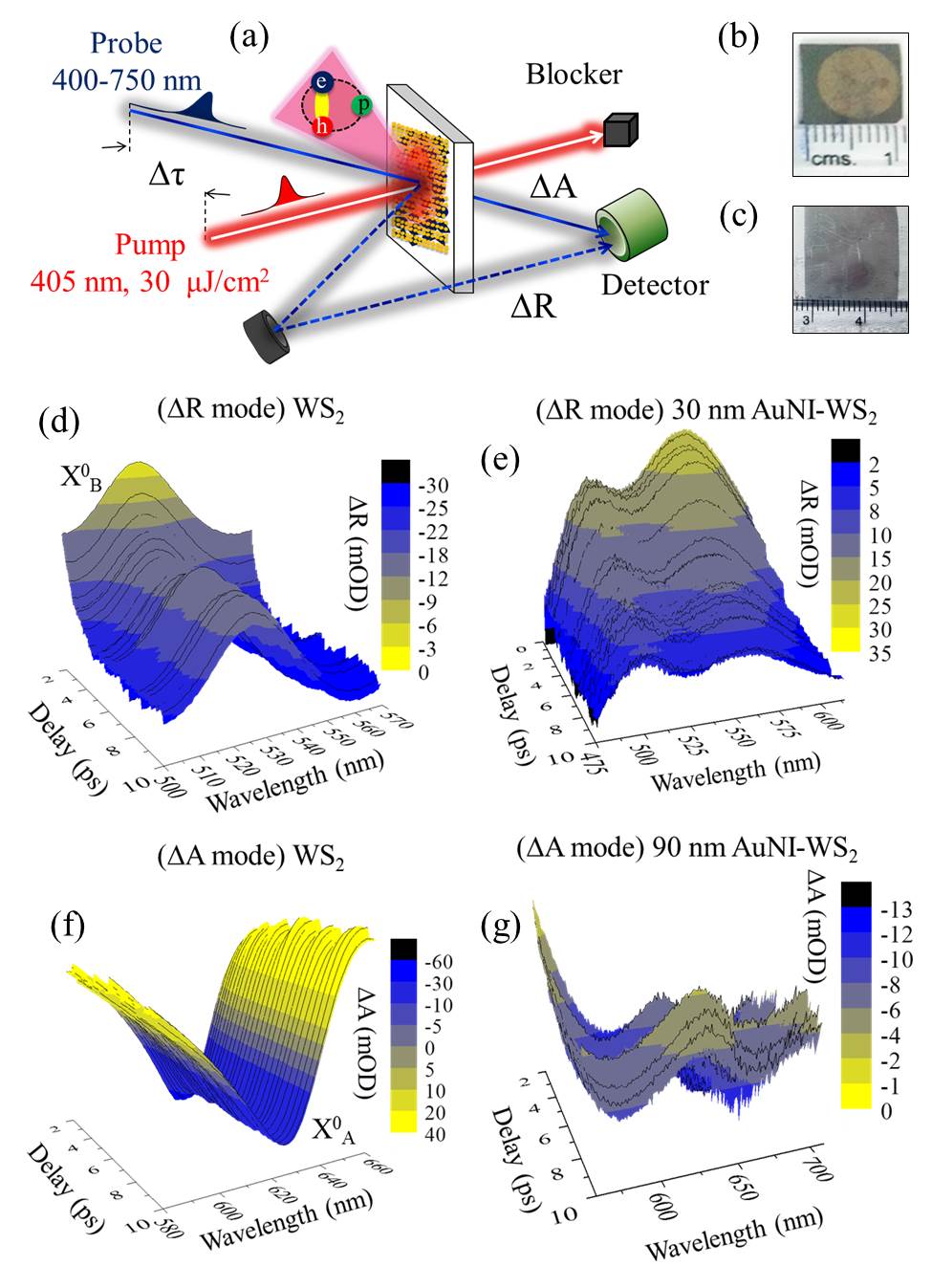}
\end{tabular}
\caption{(Color online) (a) Schematic illustration of the non-collinear transient absorption/reflection ($\Delta{A}/\Delta{R}$) setup used for the study. $\Delta\tau$ is the time delay between the pump and probe, where the probe lags behind the pump for positive delay. Figure (b-c) is the corresponding photographs of the wafer-scaled $\sim$30 nm and 90 nm $AuNI-WS_2$ hybrid samples fabricated on Si ($1{\times}1 cm^2$) and glass ($2{\times}2 cm^2$) substrates for transient measurements. Figure (d-e) and (f-g) are the corresponding $\Delta{R}$ and $\Delta{A}$ spectra of the $WS_2$ and $AuNI-WS_2$ hybrid samples fabricated on $Si$ and glass substrates, showing the transient build-up of Rabi-splittings at the respective exciton-plasmon overlap.}
\label{fig2}
\end{figure}

\subsection{Transient Rabi-splitting in Au-WS$_2$ hybrid samples}
The detailed time-resolved spin-plexciton dynamics of these fabricated nanostructures have been monitored through an ultrafast (femtosecond) helicity controlled pump-probe spectroscopy technique at room temperature (300 K), as shown in Fig.~\ref{fig2}(a). An optical parametric amplified pump beam (405 nm, $\sim{30} \mu{J}/cm^2$), generated from a $Ti$-sapphire laser (808 nm), has been used for the prior generation of both $X^0_A$ (1.97 eV) and $X^0_B$ (2.34 eV) excitons from fixed $k$-valley in layered $WS_2$ (see details in Materials and Methods). Following the pump, a time-delayed ($\Delta\tau$) broadband (1.65-2.75 eV) probe pulse has been used to record the transient absorbance ($\Delta{A} = A_{on} - A_{off}$) or reflectance ($\Delta{R} = R_{on} - R_{off}$) of $AuNI-WS_2$ hybrids fabricated on different large-scale substrates (glass for $\Delta{A}$ or $Si$ for $\Delta{R}$), as shown in Fig.~\ref{fig2}(b)-(c). Here, $A_{on}$ or $R_{on}$ and $A_{off}$ or $R_{off}$ represent the corresponding absorption or reflected probe spectrum in the presence and absence of the pump, respectively. We now discuss the time-resolved response of the as-fabricated WS$_2$ samples and $\sim$30 nm and $\sim$90 nm $AuNI-WS_2$ hybrid nanostructures by measuring $\Delta{R}$ and $\Delta{A}$, as shown in Fig.~\ref{fig2}(d)-(g). The transient spectra (Fig.~\ref{fig2}(e) and (g)) exhibit distinct transient Rabi-splitting features for both the $\sim$30 and 90 nm $AuNI-WS_2$ hybrids in comparison with the sharp excitonic peaks ($X^0_A$ and $X^0_B$) in pristine $WS_2$ (Fig.~\ref{fig2}(d) and (f)), validating the individual spin-plexciton formation. Here, both the reflection and absorption modes have been used, as in case of $\Delta{A}$, the excited state absorption of AuNI overlapped and suppressed the $X^0_B$ excitons, whereas for $\Delta{R}$, $X^0_A$ excitons are submerged due to the presence of huge pump induced reflection of $AuNI$ beyond 550 nm (Fig. S2, Supplementary Information). 

\begin{figure}[t]
\centering
\begin{tabular}{cc}
\includegraphics[width=85mm]{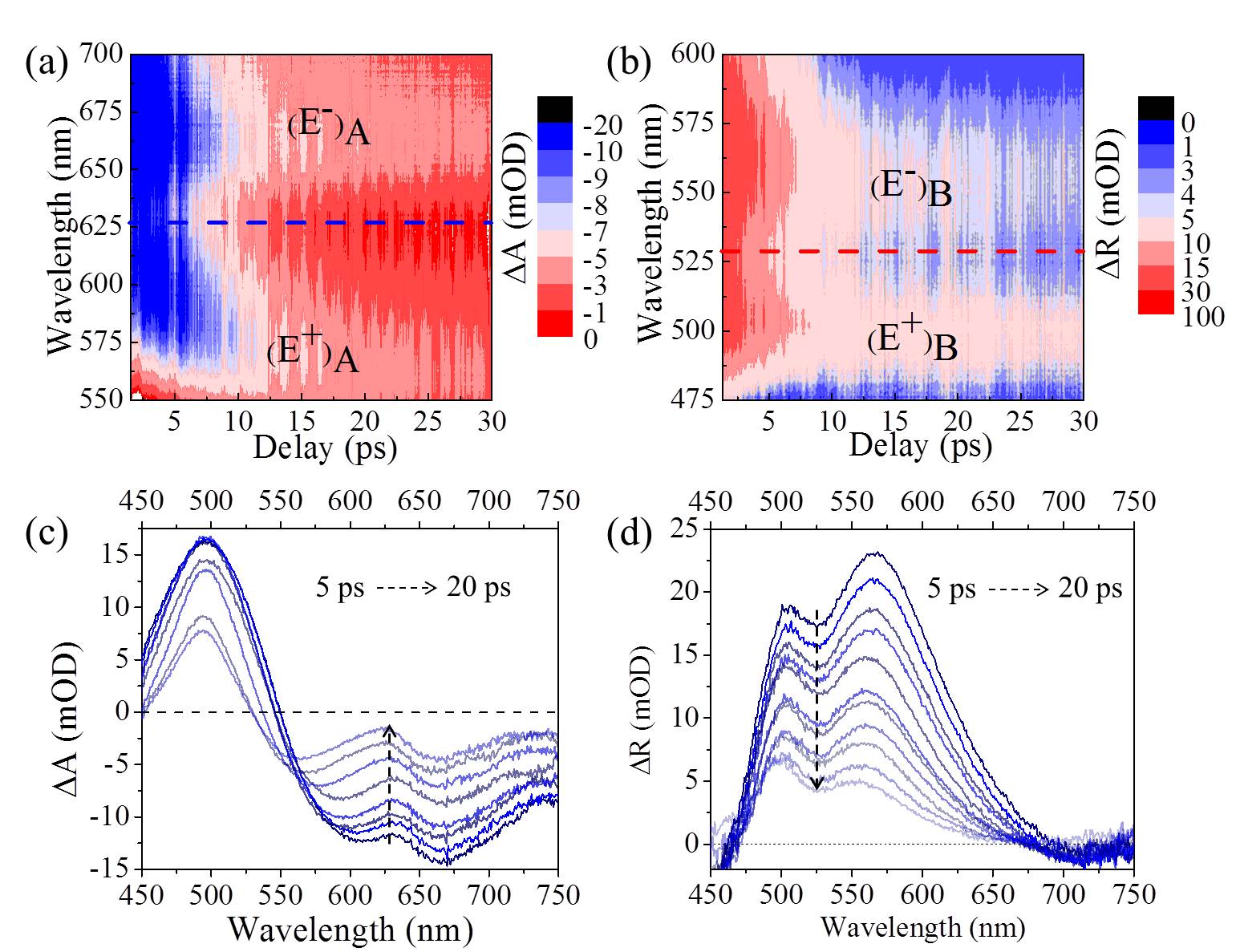}
\end{tabular}
\caption{(Color online) Contour map for (a) transient absorption of $\sim$90 nm $AuNI-WS_2$ hybrid nanostructures on glass substrate to detect $X^0_A-P$ plexcitons and (b) transient reflection of $\sim$30 nm $AuNI-WS_2$ hybrid nanostructures on $Si$ substrate to detect $X^0_B-P$ plexcitons. Both the contour maps clearly show the transient hybridized mode splitting ($E^+$ and $E^-$) for corresponding spin-plexcitons, where the position of excitons ($X^0_A$ and $X^0_B$) are denoted with blue and red dashed lines, respectively. The transient Rabi-splitting is further elaborated in figure (c-d) where individual temporal dynamics of both the plexcitons ($X^0_A-P$ and $X^0_B-P$) is shown from 5-to-20 ps.}
\label{fig3}
\end{figure}

\subsection{Spin-resolved plexcitons}
Ideally, the plexcitonic resonance should occur due to the single excitonic dipolar overlap with the plasmonic vacuum field. In our case, we have designed the metal nanostructures via dewetting of $Au$ film with variable thickness that results in self-assembled plasmonic hot-spots among $Au$ nanoislands, such that the optical field can be trapped more efficiently~\cite{r24}, as compared to an isolated single metallic absorber. Thus, these quasi-continuous plasmonic modes originated from self-assembled $AuNI$ hot-spots and pump-induced hot excitons of $WS_2$ bring the system to a classical limit, where normal hybridized mode splitting or Rabi-splitting can be realized at exciton-plasmon ($X^0-P$) resonance~\cite{r11, r25}. This splitting results in the formation of two distinct coupled polariton energy branches ($E^+$ and $E^-$) independently for both the $X^0_A-P$ and $X^0_B-P$ plexcitons, as shown in Fig.~\ref{fig3}. Previously Wang et al.~\cite{r25} suggested that the hybridized mode splitting with an energy $\Omega_R$, occurs due to the periodic energy transfer between plasmons and excitons with a time period of $\sim{{2\pi}/{\Omega_R}}$. This indicates that one may be able to track down the Rabi-oscillation with our current transient spectroscopy set-up where instrument response time is of 150 fs, only if the Rabi-splitting energy is below $\sim$150 meV. However, in our case, both the $\sim$90 and $\sim$30 nm $AuNI-WS_2$ hybrids (Fig.~\ref{fig3}(a)-(b)) achieve a transient Rabi-splitting just after the zero-delay that implies a larger Rabi-splitting energy ($\Omega_R$) value. The temporal dynamics of these hybridized modes ($E^+$ and $E^-$) and their time-resolved linewidth modulation have been investigated further as shown in Fig.~\ref{fig3}(c)-(d). Relatively broader probed ensemble spectral linewidths of the $E^+$ and $E^-$ modes point towards a higher dephasing time limit for both the plexcitons.

\begin{figure}[t]
\centering
\begin{tabular}{cc}
\includegraphics[width=85mm]{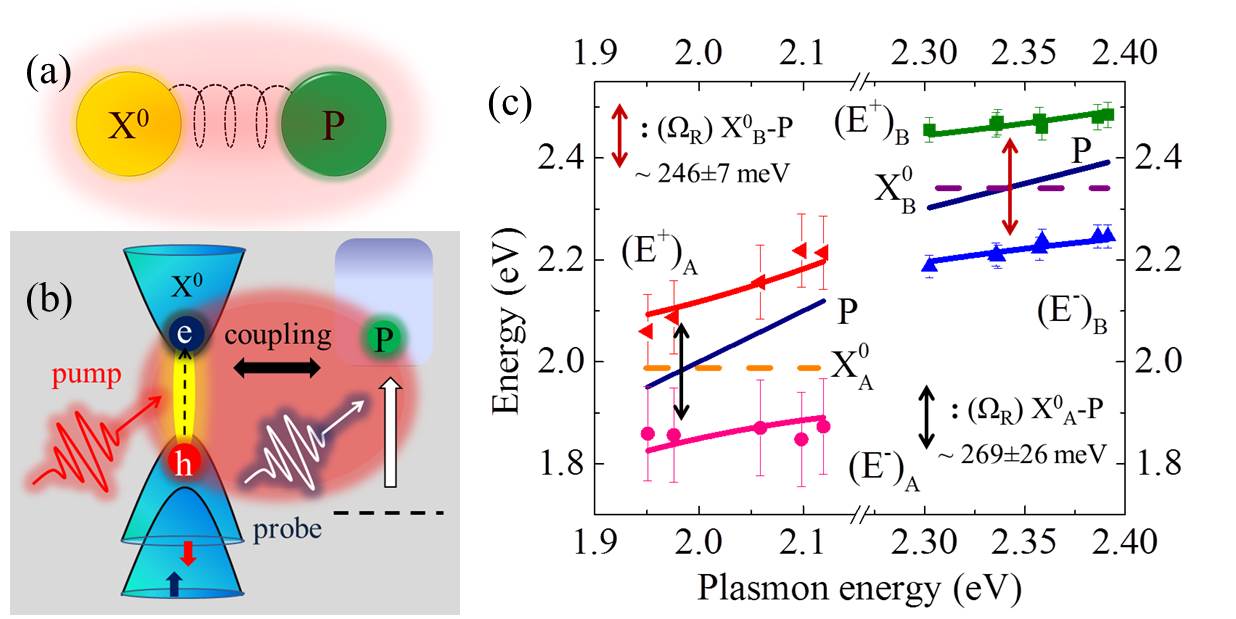}
\end{tabular}
\caption{(Color online) (a) Schematic representation of plexciton using the two-state model consisting two oscillating dipoles exciton and plasmon. The exciton (yellow sphere) and plasmon (green sphere) are coupled together through a hypothetical spring (black dashed line) which are interacting under a strong electromagnetic field. (b) An artistic representation of two-step pump-probe scheme which is equivalent to the two-state model to achieve the spin-resolved plexcitons in $AuNI-WS_2$ hybrid nanostructures. (c) Typical transient anti-crossing behavior of the hybridized energy branches ($E^+$ and $E^-$) for $X^0_A-P$ and $X^0_B-P$ plexcitons in $AuNI-WS_2$ hybrid nanostructures, as predicted in the two-state model. $X^0_A$ and $X^0_B$ are denoted as orange and purple dashed lines, whereas the navy blue line represents delay dependent blue-shifted plasmons. Individual anti-crossings have been demonstrated for both the $X^0_A-P$ and $X^0_B-P$ spin-plexcitons. The zero detuning ($\delta = 0$) or the normal mode splitting is observed at the probe delay of ~7.0 ps and the transient Rabi-splitting energies ($\Omega_R$) at zero detuning as 269$\pm$26 meV (black arrow) and 246$\pm$7 meV (red arrow) for $X^0_A-P$ and $X^0_B-P$ plexcitons, respectively.}
\label{fig4}
\end{figure} 

\subsection{Time-resolved anti-crossing}
To further investigate these peculiar transient characteristics (see details in Materials and Methods), we consider a system consisting of two oscillating dipoles ($X_0$ and $P$) with associated dipole moments ($\mu_X$ and $\mu_P$) interacting with an applied optical field, which is again coupled with one another via the resultant local electromagnetic field, as shown in Fig.~\ref{fig4}(a). In $AuNI-WS_2$ hybrids, the similar interaction between the plasmons of $AuNI$ and the excitons ($X^0_A$ and $X^0_B$) of $WS_2$ is shown in Fig.~\ref{fig4}(b), which leads to two new hybrid modes according to the two-state model~\cite{r9}. Now, the upper branch mode ($E^+$ band) and the lower branch mode ($E^-$ band) exhibit a typical anti-crossing behavior as shown in Fig.~\ref{fig4}(c), for both $X^0_A-P$ and $X^0_B-P$ plexcitons. These resultant hybrid states have both excitonic and plasmonic characters when two un-hybridized components are close to the resonance. Whereas far-off from resonance, the hybrid states are principally equivalent to the additive spectra of uncoupled $P$ and $X^0$. Here, the time-dependent anti-crossing originate due to the probe-delay dependent blue-shift of the plasmon resonance in $AuNI$~\cite{nr6, nr7, nr8} (Fig. S3, Supplementary Information). The variation in the energy amplitudes for both the branches ($E^+$ and $E^-$) is defined as~\cite{r12, r21, r24}: 

\begin{eqnarray}
E^\pm = \frac{1}{2}(E_P + E_X)\pm \sqrt{(g^2 + \frac{\delta ^2}{4})}
\label{e1}
\end{eqnarray}

where $E_P$ and $E_X$ are the corresponding energy of the plasmons ($P$) and excitons ($X^0$), $\delta = E_P - E_X$ is the detuning energy of $P$ and $X^0$, $g = \Omega_R/2 = \frac{1}{2} \sqrt{(E^+ - E^-)^2 - \delta^2}$ is the coupling rate of plexcitons. In the two-state model, we have considered both the plasmons and excitons to be ideal dipole oscillators within our considerable ultrafast time-domain ($\leq$10 ps), by neglecting the incoherent cross-damping term to simplify the model. However, the asymmetric anti-crossing behavior (Fig.~\ref{fig4}(c)) of $E^+$ and $E^-$ for both the plexcitons ($X^0_A-P$ and $X^0_B-P$) indicates the possibility of non-ideal photon exchange between $X^0$ and $P$ for these hybrids. Now, as the individual spectral positions of both the spin decoupled excitons ($X^0_A$ and $X^0_B$) are already known in $WS_2$, we can calculate the Rabi-splitting energy by directly solving $E^+$ and $E^-$. The extracted individual time-resolved Rabi-splitting energies are found to be 269$\pm$26 meV and 246$\pm$7 meV for $X^0_A-P$ and $X^0_B-P$ plexcitons, respectively, by fitting the transient peak positions of $E^+$ and $E^-$ with Eq. 1. The uncertainty value for the normal mode splitting energy is greater for $X^0_A-P$ compared to $X^0_B-P$, as this energy uncertainty is directly related with the delay-dependent broadening of the transient spectra. The size-distribution of $AuNI$ for $\sim$90 nm hybrid ($X^0_A-P$) is broader than $\sim$30 nm hybrid ($X^0_B-P$) as shown in Fig. S1(e-f) of Supplementary Information which introduces relatively higher uncertainty in the determination of spectral positions of the plexcitonic hybrid modes ($E^\pm$). Furthermore, if we consider the coupling losses in the surrounding medium, the $\Omega_R$ get modified as, $(\Omega_R)_{with  loss} \approx 2\sqrt{g^2 - \frac{{(\Delta_{X^0}-\Delta_P)}^2}{4}}$. Following this, we have extracted both the excitonic linewidth ($\Delta_{X^0}$) as $\sim$100 meV and plasmonic linewidth ($\Delta_P$) as $\sim$150 meV within $\leq$10 ps time-limit. These energies ($\Delta_{X^0}$ and $\Delta_P$) are much less than the calculated $(\Omega_R)_{with  loss}$ ($\sim$200 meV) for both $X^0_A-P$ and $X^0_B-P$ while considering the losses. This validates the strong-coupling criterion of the excitons and plasmons. Surprisingly, we discover the zero detuning (i.e. $\delta = 0$) timescale between both the excitons ($X^0_A$ and $X^0_B$) and plasmons to be close to 7.0 ps, which exactly matches with the fast decay components of both the plexcitons found experimentally and shown in Fig. S4, Supplementary Information. As the Auger scattering takes place typically in this timescale (sub-ps to few-ps) as reported earlier in the previous studies on $WS_2$~\cite{r26, nrf2}, the hot-exciton cooling process may play a crucial role to achieve this time-resolved zero detuning of both the plexcitons.

\subsection{Origin of transient Rabi-splitting}
To understand the definite origin of these robust Rabi-splitting energies, we evoke the idea of plasmonic hot-spots in $Au$ nanoislands and pump induced exciton populations in $WS_2$. In the plexcitonic limit, $E^+$ and $E^-$ can be written as $E_0 - g$ and $E_0 + g$, where $E_0 = E_P = E_X$ is the resonance energy. A detailed calculation suggests that the following condition needs to be satisfied to realize the strong coupling between the $X_0$ and $P$~\cite{r9, r11}:

\begin{eqnarray}
\Omega _R\,or\,2g \approx (\frac{D}{\omega _0}) \frac{e^2}{m_e} \mu _X \sqrt{\frac{N_X}{V_P}}
\label{e2}
\end{eqnarray}

where ${D}$ is the dipolar coupling factor $\propto$ 1/(diameter of $AuNI$)$^3$, ${\omega_0}$ is the resonance frequency, ${e}$ is electronic charge, $\it{m_e}$ is the effective mass of the electron, ${\mu_X}$ is exciton dipole moment, ${N_X}$ is the effective number of excitons in the system and ${V_P}$ is the plasmon mode volume. During our transient measurements, the pump fluence has been maintained around $\sim$30 $\mu$J/cm$^2$ with a pump photon energy of 3 eV throughout the experiments to avoid any pump induced many-body effects and spectral linewidth modulation due to hot-exciton population in $WS_2$~\cite{r26}. Thus, the effective number of both the excitons ($X^0_A$ and $X^0_B$) in the system always reaches to a certain population right before the occurrence of probe excitation. On the other hand, plasmonic hot-spots can trap sufficient number of probe-photons to create quasi-continuum of plasmons. As a result of this, both the plexcitons ($X^0_A-P$ and $X^0_B-P$) can be accomplished facilely via strong dipolar coupling among pre-excited pump-induced excitons and probe-induced plasmons. Thus, we conclude that our system reaches a temporally stable strong coupling regime at a probe delay of $\sim$7.0 ps, where the normal mode splitting energy exceeds $\sim$250 meV for both the spin decoupled plexcitons.

\noindent
\section{Conclusions}
\label{summary}
\noindent
In summary, our report reveals individual time-resolved generation and detection of both the spin-locked plexcitons in metal-TMDs hybrids. We attribute strong coupling condition between plasmons and spin-resolved excitons in size-tunable $AuNI-WS_2$ hybrid nanostructures through a two-step excitation process that displays clear transient anti-crossing behavior for both the spin-resolved plexcitons, in agreement with the results of two-state model. A remarkably robust Rabi-splitting energy ($\sim$250 meV) and comparatively higher zero detuning time ($\sim$7.0 ps) are realized for both the spin-resolved plexcitons of $AuNI-WS_2$ hybrids. We thereby strongly believe that our results suggest new possibilities to study the ultrafast behavior of the spin-plexcitons in time-domain and may be an interesting route to explore possible spin-polaritonic device aspects at room temperature in future.  

\section{Acknowledgments}

\noindent
Authors acknowledge SDGRI-UPM project of IIT Kharagpur for necessary equipment support in Ultrafast Science Lab, Department of Physics, IIT Kharagpur. RKC thanks Sayantan Bhattacharya and Manobina Karmakar for their inputs during transient measurements.

\section{Appendix: Materials and Methods}
\subsection{Sample fabrication}
\noindent
Quasi-continuous layered $WS_2$ films (typically $\sim$10 nm thick) were fabricated using freshly exfoliated mono-to-few layer $WS_2$ (Sigma-Aldrich) ink homogeneously dispersed in dimethylformamide ($DMF$, 99.9$\%$, Sigma-Aldrich) as reported earlier in our previous work~\cite{r27}, and spin-coated at 1,500 rotation per minute onto glass (2$\times$2 cm$^2$) or silicon (1$\times$1 cm$^2$) substrates at 90 ${^0}C$ under an inert condition. Thereafter, optically thick gold films with variable thickness (controlled through precursor percentage) were deposited on as-fabricated $WS_2$ layers using thermal evaporation at $\sim$10$^{-6}$ mbar chamber pressure. Following this, the samples were annealed for 3 hours at 300 $^{0}C$ constant temperature under an inert ($Ar$) atmosphere, resulting in the desired size of gold nanoislands ($AuNI)-WS_2$ hybrid nanostructures as shown via the step-by-step fabrication schematic in Fig. S5 of the Supplementary Information. The fabricated samples can act as wafer-scaled emergent prototypes for exploring the transient dynamics of spin-resolved plexcitonic modes. The sample fabricated on glass ($\sim$90 nm $AuNI-WS_2$) was used for the transient absorption and that on a $Si$ substrate ($\sim$30 nm $AuNI-WS_2$) was used for transient reflection measurements.

\subsection{Transient measurements set-up}
\noindent
Ultrafast transient absorption/reflection spectra of $AuNI-WS_2$ on glass/silicon was recorded with a commercially available transient spectrometer (Newport) using 50 fs $Ti$: Sapphire mode locked amplifier (Libra-He, Coherent) with 808 nm center wavelength and a repetition rate of 1 kHz with 3 mJ average pulse energy. Principle part (70$\%$) of the fundamental beam (808 nm) was fed into an optical parametric amplifier (TOPAS prime, Coherent), which generated horizontally polarized 405 nm ($\sim$3 eV) pump beam that converts into right circularly polarized output by passing through a broadband (400-800 nm) achromatic quarter-wave plate (Thorlabs). The minor part (30$\%$) of the fundamental beam was focused on to a $CaF_2$ crystal, which generated a stable white light continuum (400-750 nm) as a probe spectrum. This probe spectrum passed through an optical delay channel using a motorized translational stage (ILS-LM, Newport) in order to probe the system at different times, after the pump excitation was employed to perturb the system. Pump and probe beam spot-sizes were maintained at 1.0 mm and $\sim$100 $\mu$m, respectively which were used for our non-collinear pump-probe measurement setup. This kind of arrangement provided a pump energy fluence of $\sim$30 ${\mu}$J/cm$^{2}$ with the pump:probe fluence ratio of 300:1. A spectrometer equipped with a fiber coupled linear $Si$ photodiode array (MS-260i, Oriel Instrument) was used to detect the differential probe spectra. Differential absorbance ($\Delta{A}$)/reflectance ($\Delta{R}$) spectra were measured by modulating the pump beam at 500 Hz with an optical chopper, which was ultimately fed to the spectrometer. Here, this $\Delta{A}$ or $\Delta{R}$ are measured in terms of $mOD$ which is the change in optical density$\times$10$^{-3}$. 

\subsection{Two-state model}
\noindent
In our $AuNI-WS{_2}$ hybrid nanostructures, self-assembled $Au$ nanoislands produce strong hot-spots on top of $WS{_2}$ layers. As a result, only a small fraction of $WS_2$ experiences a strong near-field coupling with quasi-continuous plasmonic hot-spots, which results in the formation of plexcitons.

Following Thomas et al.~\cite{r9}, we have considered the coupled harmonic oscillator model where we assume two dipoles (one associated with excitons ($X^0$) and the other one for plasmons ($P$)) interacting strongly under an applied local electromagnetic field (EM) (as shown in Fig.~\ref{fig4}(a)). The equation of interaction can be written as~\cite{r9, r24},

\begin{multline}
\ddot \mu_X + {\gamma}_{X} \dot \mu_X + \omega ^2_X \mu_X = A_X [Ecos(\omega t) + D \mu_P]
\end{multline} 
\begin{multline}
\ddot \mu_P + {\gamma}_{P} \dot \mu_P + \omega ^2_P \mu_P = A_P [Ecos(\omega t) + D \mu_X]
\end{multline}

where $\mu_X$ and $\mu_P$ are the dipole moments of the exciton and plasmon respectively, $\omega_i$ ($= E_i / \hbar$) are the corresponding resonance frequency, $A_i$ is related with the oscillator strengths, $\gamma_i$ the damping constant of the oscillators (here, $i = X$ and $P$), $Ecos(\omega t)$ the driving local EM field and $D \propto 1/r^3$, where r is the center-to-center distance between dipoles. Solving these two equations, we can calculate both the dipole moments ($\mu_X$ and $\mu_P$) that is useful to calculate the absorption cross-sections ($\sigma$) and exciton-plasmon coupling energy ($g$)~\cite{r9, rnew1}.

Now, to calculate the hybrid mode energies of the upper and lower branches ($E^\pm$), two-state model is incorporated, as shown in Fig.~\ref{fig4}(b). $g (= \Omega_R/2$) is the exciton-plasmon coupling energy which is essential to achieve the strong-coupling condition of plexcitons. Now, one can write the dipolar interaction in terms of eigen value equation,
\begin{eqnarray}
H\mu_i = \begin{bmatrix} E_X & g \\ g & E_P \end{bmatrix} \mu_i
\label{e5}
\end{eqnarray}
Thus, the eigen values of the Hamiltonian matrix is equivalent to the hybrid mode energies of the plexcitonic system ($E^\pm$) as shown in the Eq. 1. 
In Eq. 5, we have not considered the damping term ($\gamma_i$) that will modify the Eq. 1 as~\cite{r9},
\begin{multline}
E^\pm = \frac{1}{2}(E_P + E_X)\pm \sqrt{g^2 + \frac{\delta ^2}{4} - \frac{{\hbar}^2}{16} (\gamma_P - \gamma_X)^2} \\
 - j\frac{\hbar}{4}(\gamma_P + \gamma_X)
\end{multline}
when we replace $E_i \rightarrow E_i - j \frac{\hbar \gamma_i}{2}$,($i = X$ and $P$, $j = \sqrt{-1}$) in Eq. 5. As our system validates the strong coupling conditions of plexcitons, i.e., $g \gg \frac{\hbar \mid \gamma_P - \gamma_X \mid}{4}$, Eq. 6 will thus reduce to Eq. 1.

\end{document}